\documentclass[aps,prd,twocolumn,showpacs,floats,floatfix,letterpaper,footinbib,superscriptaddress,]{revtex4}

\usepackage{amssymb,amsmath,latexsym,mathrsfs}
\usepackage{graphicx}
\usepackage{epsfig}
\usepackage{multirow}
\usepackage{array}
\usepackage{color}

\newcommand{\ueV}{\mu\mathrm{eV}}
\newcommand{\eV}{\mathrm{eV}}

\newcommand{\MeV}{\mathrm{MeV}}
\newcommand{\GeV}{\mathrm{GeV}}
\newcommand{\Mpc}{\mathrm{Mpc}}
\newcommand{\neff}{N_\mathrm{eff}}

\begin{document}


\title{Axion cold dark matter: status after Planck and BICEP2}
\author{Eleonora Di Valentino}
\affiliation{Physics Department and INFN, Universit\`a di Roma ``La Sapienza'', Ple Aldo Moro 2, 00185, Rome, Italy}
\author{Elena Giusarma}
\affiliation{Physics Department and INFN, Universit\`a di Roma ``La Sapienza'', Ple Aldo Moro 2, 00185, Rome, Italy}
\author{Massimiliano Lattanzi}
\affiliation{Dipartimento di Fisica e Science della Terra, Universit\`a di Ferrara and INFN,\\
sezione di Ferrara, Polo Scientifico e Tecnologico - Edficio C Via Saragat, 1, I-44122 Ferrara Italy}
\author{Alessandro Melchiorri}
\affiliation{Physics Department and INFN, Universit\`a di Roma ``La Sapienza'', Ple Aldo Moro 2, 00185, Rome, Italy}
\author{Olga Mena}
\affiliation{IFIC, Universidad de Valencia-CSIC, 46071, Valencia, Spain}

\begin{abstract}

We investigate the axion dark matter scenario (ADM), in which axions account for all of the dark matter in the Universe,
in light of the most recent cosmological data. In particular, we use the Planck temperature data, complemented by WMAP
E-polarization measurements, as well as the recent BICEP2 observations of
B-modes. Baryon Acoustic Oscillation data, including those from the Baryon Oscillation
Spectroscopic Survey, are also considered in the numerical analyses.

We find that, in the minimal ADM scenario, the full dataset implies that the axion mass $m_{\textrm{a}} = 82.2 \pm 1.1\,\mu\eV$
(corresponding to the Peccei-Quinn symmetry being broken at a scale $f_{\textrm{a}} = (7.54 \pm 0.10) \times 10^{10} \,\GeV$), or
 $m_{\textrm{a}} = 76.6 \pm 2.6\,\mu\eV$ ($f_{\textrm{a}} = (8.08 \pm 0.27) \times 10^{10} \,\GeV$)
when we allow for a non-standard effective number of relativistic species $\neff$.
We also find a 2$\sigma$ preference for $\neff>3.046$. The limit on the sum of
neutrino masses is
$\sum m_\nu < 0.25\,\eV$ at 95\% CL for $\neff=3.046$, or $\sum
m_\nu<0.47\,\eV$ when $\neff$ is a free parameter.

Considering extended scenarios where either the dark energy equation-of-state parameter $w$,
the tensor spectral index $n_t$ or the running of the scalar index $d n_s/d\ln k$ are allowed to vary
does not change significantly the axion mass-energy density
constraints. However, in the case of the full dataset exploited here, there is a preference for a non-zero tensor index or scalar running, driven by the different tensor amplitudes
implied by the Planck and BICEP2 observations.

Dark matter axions with mass in the $70 - 80\,\ueV$ range can, in principle, be detected by looking for axion-to-photon
conversion occurring inside a tunable microwave 
cavity permeated by a high-intensity magnetic field, and operating at a frequency $\nu \simeq 20$ GHz. This is out of the
reach of current experiments like ADMX (limited to a maximum frequency of a few GHzs), but is, on the other hand, 
within the reach of the upcoming ADMX-HF
experiment, that will explore the $4 - 40$ GHz frequency range and then being sensitive to axions masses up to $\sim 150 \,\ueV$.

\end{abstract}


\maketitle

\section{Introduction}

In the past few years, a great deal of observational evidence has accumulated
in support of the $\Lambda$CDM cosmological model, that is to date regarded as the 
"standard model" of cosmology. One of the great puzzles related to the $\Lambda$CDM model,
however, is the nature of the dark matter (DM) component that, according to the recent
Planck observations~\cite{planck,Ade:2013ktc,Planck:2013kta}, makes up roughly 27\% of the total matter-energy content of the Universe.
A well-motivated DM candidate is the axion, that was first proposed by Peccei and Quinn \cite{PecQui77}
to explain the strong CP problem, i.e., the absence of CP violation in strong interactions.

Here we consider the hypothesis that the axion accounts for all the DM present in the Universe.
We put this "axion dark matter" (ADM) scenario under scrutiny using the most recent cosmological data,
in particular the observations of cosmic microwave background (CMB)
temperature~\cite{planck,Ade:2013ktc,Planck:2013kta} and 
polarization anisotropies (including the recent BICEP2 detection of B-mode polarization
\cite{Ade:2014xna,Ade:2014gua}) and of Baryon Acoustic Oscillations
(BAO)~\cite{Anderson:2013vga,dr71,dr72,wigglez,6df}. 
The ADM model has also been revisited by other authors~\cite{Visinelli:2014twa,Marsh:2014qoa} in light of
BICEP2 data, and our analyses in the minimal $\Lambda$CDM scenario agree with these
previous works. However, in order to assess the robustness of the cosmological
constraints presented in the literature, as well as the tension between BICEP2
and Planck measurements of the tensor-to scalar ratio, here we also consider extensions of the simplest ADM
model. The effects of additional relativistic degrees of freedom,
of neutrino masses, of a dark energy equation-of-state
parameter and of a free-tensor spectral index are carefully explored.

The paper is structured as follows. Section~\ref{sec:sec1} introduces
axions in cosmology and derives the excluded scenarios after BICEP2
data. Section~\ref{sec:sec2} describes the analysis method and the data
used to analyse the ADM  models that survive after applying BICEP2 
bounds on gravitational waves and Planck isocurvature constraints.
In Sec.~\ref{sec:sec3} we present our main results and we draw our
conclusions in Sec.~\ref{sec:sec4}.

\section{Axion cosmology}
\label{sec:sec1}
In order to solve the strong CP problem dynamically, Peccei and Quinn postulated
the existence of a new global $U(1)$ (quasi-) symmetry, often denoted $U(1)_{\mathrm{PQ}}$,
that is spontaneously broken at the Peccei-Quinn (PQ) scale $f_{\textrm{a}}$. The spontaneous
breaking of the PQ symmetry generates a pseudo Nambu-Goldstone boson,
the axion, which can be copiously produced in the universe's early
stages, both via thermal and non-thermal processes. Thermal axions with sub-eV masses
contribute to the hot dark matter component of the universe, as
neutrinos, and the cosmological limits on their properties have
been recently updated and presented in
Refs.~\cite{Archidiacono:2013cha,Giusarma:2014zza}.

Here we focus on axion-like particles produced non-thermally, as they
were postulated as natural candidates for the
cold dark matter component~\cite{Preskill:1982cy,Abbott:1982af,Dine:1982ah,Turner:1990uz,Lyth:1992tw}.
The history of axions start at the PQ scale $f_{\textrm{a}}$. For temperatures
between this scale and the QCD phase transition $\Lambda_{\textrm{QCD}}$, the axion is, for
practical purposes, a massless particle. When the universe's temperature
approaches $\Lambda_{\textrm{QCD}}$, the axion acquires a mass via
instanton effects. The effective potential $V$ for the axion field $a(x)$ is generated through 
non-pertubative QCD effects \cite{Gro80} and, setting the color
anomaly $N=1$, it may be written as
\begin{equation}
V(a)=f_{\textrm{a}}^2 m_{\textrm{a}}^2(T)\left[1-\cos\left(\frac{a}{f_{\textrm{a}}}\right)\right] \,,
\label{eq:potential}
\end{equation}
where the axion mass is a function of temperature. Introducing the misalignment angle
$\theta \equiv a/f_{\textrm{a}}$, the field evolves according to the Klein-Gordon equation on a flat Friedmann-Robertson-Walker
background:
\begin{equation}
\ddot \theta + 3 H\dot \theta+m_{\textrm{a}}^2(T)\theta= 0~,
\label{eq:aprox}
\end{equation}
where the axion temperature-dependent mass is \cite{Gro80}
\begin{equation}
m_{\textrm{a}}(T)=\left\{
\begin{array}{lll}
&C m_{\textrm{a}} (T=0)  (\Lambda_\mathrm{QCD}/T)^4 \quad &T \gtrsim \Lambda_\mathrm{QCD}\\[0.2cm]
&m_{\textrm{a}} (T=0) \quad & T \lesssim \Lambda_\mathrm{QCD}
\end{array}
\right.
\label{eq:maT1}
\end{equation}
where $C\simeq 0.018$ is a model dependent factor, see Refs~\cite{Gro80,Fox:2004kb}, $\Lambda_{\mathrm{QCD}}\simeq 200\,\MeV$ and the zero-temperature
 mass $m_{\textrm{a}} (T=0)$ is related to the PQ scale:
\begin{equation}
m_{\textrm{a}} \simeq 6.2\,\ueV \left(\frac{f_{\textrm{a}}}{10^{12}\,\GeV}\right)^{-1} ~.
\end{equation}
The axion is effectively massless at $T\gg \Lambda_\mathrm{QCD}$, as
it can be seen from Eq. (\ref{eq:maT1}).

The PQ symmetry breaking can occur before or after inflation. If there
was an inflationary period in the universe after or during the PQ phase
transition, there will exist, together with the standard adiabatic
perturbations generated by the inflaton field, axion isocurvature
perturbations, associated to quantum fluctuations in the axion field. In this scenario, i.e.
when the condition
\begin{equation}
f_{\textrm{a}} >\left(\frac{H_I}{2\pi}\right)~,
\end{equation}
is satisfied, the initial misalignment angle $\theta_i$ should be identical in the whole
observable universe, with a variance given by 
\begin{equation}
\langle\sigma_\theta^2 \rangle = \left(\frac{H_I}{2\pi f_{\textrm{a}}}\right)^2~,
\end{equation}
and corresponding to quantum fluctuations in the massless axion field 
\begin{equation}
\langle\delta_a^2 \rangle = \left(\frac{H_I}{2\pi}\right)^2~,
\end{equation}
\newline
\noindent
where $H_I$ is the value of the Hubble parameter during
inflation. These quantum fluctuations generate
an axion isocurvature power spectrum
\begin{equation}
\Delta_a (k)=  k^3 |\delta_a^2|/2 \pi^2= \frac{H_I^2}{\pi^2}\theta_i^2
f_{\textrm{a}}^2~.
\end{equation}
The Planck data, combined with the 9-year polarization data from WMAP
\cite{Bennett:2012fp} constrain the primordial isocurvature fraction (defined as the
ratio of the isocurvature perturbation spectrum to the sum of 
the adiabatic and isocurvature spectra) to be~\cite{Ade:2013uln}
\begin{equation}
\beta_{\textrm{iso}} < 0.039~,
\end{equation}
at $95\%$~CL and at a scale $k=0.05$~Mpc$^{-1}$. This limit can be
used to exclude regions in the parameter space of the PQ scale  and
the scale of inflation $H_I$, since they are related via 
 \begin{widetext}
\begin{equation}
H_I=0.96\times 10^7 \
\textrm{GeV}\left(\frac{\beta_{\textrm{iso}}}{0.04}\right)^{1/2}
\left(\frac{\Omega_{\textrm{a}}}{0.120}\right)^{1/2}\left(\frac{f_{\textrm{a}}}{10^{11} \ \textrm{GeV}}\right)^{0.408}~,
\end{equation}
 \end{widetext}
where $\Omega_{\textrm{a}}$ is the axion mass-energy density.
 In this scenario, in which the PQ
symmetry is not restored after inflation, and therefore the condition 
 $f_{\textrm{a}} >\left(\frac{H_I}{2\pi}\right)$ holds, and assuming that the dark matter
 is made of axions produced by the misalignment mechanism~\footnote{An
   additional relevant mechanism of axion production that we will
   shortly see is via the decay of axionic strings. However, in this
   particular scenario such contribution will be negligible, since
   strings and other defects are diluted after the inflationary
   stage.}, Planck data has set a $95\%$~CL upper bound on the energy scale of 
 inflation~\cite{Ade:2013uln}
\begin{equation}
H_I \le 0.87\times 10^7 \	
\textrm{GeV}\left(\frac{f_{\textrm{a}}}{10^{11} \ \textrm{GeV}}\right)^{0.408}~.
\end{equation}
Very recently the BICEP2 collaboration has reported $6\sigma$ evidence for the
detection of primordial gravitational waves, with a tensor to scalar
ratio $r=0.2^{+0.07}_{-0.05}$, pointing to inflationary energy scales
of $H_I \sim 10^{14}$~ GeV
\cite{Ade:2014xna,Ade:2014gua}. These scale would require
a value for $f_{\textrm{a}}$ which lies several orders of magnitude above the Planck
scale and consequently nullifies the axion scenario in which the PQ is
broken during inflation. We conclude that, if future CMB polarisation
experiments confirm the BICEP2 findings, the axion scenario in which
the PQ symmetry is broken during inflation will be ruled out, at least in its simplest form. 
This conclusion could be circumvented in a more complicated scenario (see e.g. Ref. \cite{Higaki:2014ooa} for 
a proposal in this direction) but we shall not consider this possibility here.

There exists however another possible scenario in which the PQ
symmetry is broken after inflation, i.e
\begin{equation}
f_{\textrm{a}} <\left(\frac{H_I}{2\pi}\right)~,
\end{equation}
In this second axion cold dark matter scheme, there are no axion isocurvature perturbations since there are not axion
quantum fluctuations. On the other hand, there
will exist a contribution to the total axion energy density from
axionic string decays. We briefly summarise these two contributions
(misalignment and axionic string decays) 
to $\Omega_{\textrm{a}} h^2$. The misalignment mechanism will produce an initial
axion number density which reads
\begin{equation}
n_a(T_1)\simeq \frac{1}{2}m_{\textrm{a}}(T_1) f_{\textrm{a}}^2 \theta_i^2 
\end{equation}
where $T_1$ is defined as the temperature for which the condition $m_{\textrm{a}}(T_1)=3H(T_1)$ 
is satisfied. The mass-energy density of axions today related to
misalignment production is obtained via the product of the
ratio of the initial axion number density to entropy density times the
present entropy density, times the axion mass $m_{\textrm{a}}$, and reads~\cite{Visinelli:2009zm}
  
\begin{equation}
\Omega_{\textrm{a,mis}}h^2=\left\{
\begin{array}{lll}
&0.236\langle \theta_i^2f(\theta_i)\rangle \left(\frac{f_{\textrm{a}}}{10^{12}\ \textrm{GeV}}\right)^{7/6}  \quad &f \lesssim \hat{f}_a\\[0.2cm]
&0.0051\langle \theta_i^2f(\theta_i)\rangle \left(\frac{f_{\textrm{a}}}{10^{12}\ \textrm{GeV}}\right)^{3/2}  \quad &f \gtrsim \hat{f}_a
\end{array}
\right.
\label{eq:maT}
\end{equation}
where $\hat{f}_a=9.91\times 10^{16}$~GeV and $f(\theta_i)$ is a function related to
anharmonic effects, linked to the
fact that Eq.~(\ref{eq:aprox}) has been obtained assuming that the
potential, Eq.~(\ref{eq:potential}), is harmonic. The value of
$\theta_i^2$ is an average of a uniform distribution of all possible initial
values:
\begin{equation}
\langle \theta_i^2f(\theta_i)\rangle=\frac{1}{2 \pi} \int_{-\pi}^{\pi}
\theta_i^2f(\theta_i) d\theta_i~.
\end{equation}
If we now consider the recent BICEP2 results, the value of $f_{\textrm{a}}$,
which, in this second scenario, should be always smaller than the
inflationary energy scale, will always be smaller than $\hat{f}_a$ and
therefore, the misalignment axion cold dark matter energy density is
\begin{equation}
\Omega_{\textrm{a,mis}}h^2 = 2.07 \left(\frac{f_{\textrm{a}}}{10^{12}\
    \textrm{GeV}}\right)^{7/6}~.
\end{equation}
As previously stated, there will also be a contribution from axionic
string decays, $\Omega_{\textrm{a,dec}}h^2$.  The total (axion)
cold dark matter density $\Omega_{\textrm{a}}h^2$ is the sum of the misalignment and string decay contributions ~\cite{Visinelli:2014twa}:

\begin{multline}
\Omega_{\textrm{a}}h^2=2.07\,(1+\alpha_{\textrm{dec}}) \left(\frac{f_{\textrm{a}}}{10^{12}\
    \textrm{GeV}}\right)^{7/6}  \ ,
\label{eq:axionomega}
\end{multline}
where $\alpha_{\textrm{dec}}$ is the ratio $\alpha_{\textrm{dec}}=\Omega_\mathrm{a,dec}/ \Omega_\mathrm{a,mis}$ between the two contributions.\\
Following Ref.~\cite{Visinelli:2009zm}, we consider $\alpha_{\textrm{dec}}=0.164$ 
so that\footnote{A value for the axionic string decays fractional contribution larger than 
the one used here has been recently reported in Ref. ~\cite{beck}. 
This value is obtained by combining the observed value of $\Omega_{\mathrm{c}} h^2$
 with estimates of the axion mass based on the Josephson effect.}

\begin{equation}
\Omega_{\textrm{a}}h^2=2.41 \left(\frac{f_{\textrm{a}}}{10^{12}\textrm{GeV}}\right)^{7/6}  \, .
\end{equation}

In the following we will quote our results on $m_a$ for the case $\alpha_{\textrm{dec}}=0.164$.
However, the CMB is actually only sensitive to $\Omega_{\textrm{a}}h^2 \propto (1+
\alpha_\mathrm{dec}) m_a^{-7/6}$, therefore limits on $m_a$ for an arbitrary value of $\alpha_{\textrm{dec}}$
can be obtained from the ones reported in the following section by means of the rescaling:
\begin{equation}
m_a \longrightarrow m'_{\textrm{a}}=m_{\textrm{a}} \left[\frac{(1+\alpha_{\textrm{dec}})}{(1+0.164)}\right]^{6/7}  \ .
\end{equation}

\section{Method}
\label{sec:sec2}

The basic ADM scenario analysed here is
described by the following set of parameters:
\begin{equation}
\label{parameter}
  \{\omega_{\textrm{b}},\,\theta_s,\,\tau,\,n_s, \log[10^{10}A_{s}],\,r,\,m_{\textrm{a}}\} \, ,
\end{equation}
where $\omega_{\textrm{b}}\equiv\Omega_{\textrm{b}}h^{2}$ is the physical baryon density,
$\theta_{s}$ the ratio of the sound horizon to the angular
diameter distance at decoupling, $\tau$ is the reionization optical depth,
$A_s$ and $n_s$ are, respectively, the amplitude and spectral index of the
primordial spectrum of scalar perturbations, $r$ is the ratio between the amplitude
of the spectra of tensor and scalar perturbations, and finally $m_{\textrm{a}}$ is the axion mass. 
The latter sets the density of cold dark matter $\Omega_{\textrm{c}} h^2\equiv\Omega_{\textrm{a}} h^2$ through
Eq.~(\ref{eq:axionomega}). All the quantities characterising the primordial scalar and tensor spectra (amplitudes, spectral indices, possibly running)
are evaluated at the pivot wave number $k_0 = 0.05\,\Mpc^{-1}$.
In the baseline model we assume flatness, purely adiabatic initial conditions, a total neutrino mass $\sum m_\nu = 0.06 \,\eV$
and a cosmological constant-like dark energy ($w=-1$). We also assume,
unless otherwise noted, that the inflation consistency condition $n_T = -r/8 $ 
between the tensor amplitude and spectral index holds.

Extensions to the baseline model described above are also explored. We start by considering the effective number
of relativistic degrees of freedom and the sum of neutrino masses, first separately 
and then jointly, as additional parameters. A model with $\Delta \neff$
sterile massive neutrino species, characterised by a sterile neutrino mass 
$m^\textrm{eff}_s$, is also analysed.  Then we proceed to add the equation-of-state parameter $w$
of dark energy to the baseline model. Finally, we also study the effect 
of having more freedom in the inflationary sector, by letting the
running of the scalar spectral index vary or by relaxing the
assumption of  the inflation consistency.

We use the CAMB Boltzmann
code~\cite{camb} to evolve the background and perturbation equations,
and derive posterior distributions for the model parameters from current data
using a Monte Carlo Markov Chain (MCMC) analysis based on the publicly
available MCMC package \texttt{cosmomc}~\cite{Lewis:2002ah} that implements
the Metropolis-Hastings algorithm.

\subsection{Cosmological data}

We consider the data on CMB temperature anisotropies measured 
by the Planck satellite
 \cite{Ade:2013ktc,Planck:2013kta}
supplemented by the 9-year polarization data from WMAP \cite{Bennett:2012fp}.
   
The likelihood functions associated to these datasets are estimated 
and combined using the likelihood code
distributed by the Planck collaboration, described in Ref. \cite{Planck:2013kta}, and publicly
available at Planck Legacy Archive\footnote{\protect\url{http://pla.esac.esa.int/pla/aio/planckProducts.html}}. 
We use Planck TT data up to a maximum multipole number of $\ell_{\rm max}=2500$, 
and WMAP 9-year polarization data  (WP) up to $\ell=23$
\cite{Bennett:2012fp}.

Very recently, the BICEP2 collaboration has reported evidence  for the
detection of B-modes in the multipole range
$30 <\ell< 150$ after three
seasons of data taking in the South
Pole~\cite{Ade:2014xna,Ade:2014gua} with $6\sigma$ significance.
This B-mode excess is much higher than
known systematics and expected foregrounds, being the spectrum well fitted with a
tensor-to-scalar ratio $r=0.2^{+0.07}_{-0.05}$. Notice however
that when foregrounds are taken into account, 
subtraction of different foreground models makes the best-fit value for $r$ move
in the range $0.12 - 0.21$. In the following we shall nevertheless assume
that the BICEP2 signal is entirely of cosmological origin. The likelihood data
from the BICEP2 collaboration has been included in our MCMC analyses
accordingly to the latest version of the \texttt{cosmomc} package.

We also use BAO measurements, namely the SDSS Data
Release 7~\cite{dr71,dr72}, WiggleZ survey~\cite{wigglez} and 
6dF~\cite{6df} datasets, as well as the most recent and most accurate BAO
measurements to date, arising from the BOSS Data Release 11 (DR11)
results~\cite{Anderson:2013vga}.

\section{Results}
\label{sec:sec3}
\begin{table*}
\begin{center}
\resizebox {19cm}{!}{
\begin{tabular}{|c|c|c|c|c|c|c|c|c|}
\hline\hline
Parameter & ADM+r & ADM+r & ADM+r & ADM+r & ADM+r & ADM+r & ADM+r & ADM+r \\
&&+$N_{\textrm{eff}}$&+$\sum m_{\nu}$&+$\sum
m_{\nu}$+$N_{\textrm{eff}}$& + $m^\textrm{eff}_s+
N_{\textrm{eff}}$ & +$w$ &+ $n_t$ & + $d  n_s/d\ln k$\\
\hline
$\Omega_{\textrm{b}}h^2$ &$0.02204\pm0.00028$ &$0.02261\pm0.00043$
&$0.02189\pm0.00033$ &$0.02245\pm0.00047$   &
$0.02246\pm0.00039$ &$0.02208\pm0.00028$ &$0.02211\pm0.00029$
&$0.02229\pm0.00031$\\
$\Omega_{\textrm{a}}h^2$ &$0.1194\pm0.0027$ &$0.1280\pm0.0054$
&$0.1203\pm0.0029$ &$0.1277\pm0.0054$ & $0.1275\pm0.0055$ 
&$0.1192\pm0.0026$ &$0.1206\pm0.0030$ &$0.1198\pm0.0027$\\
$\theta$ &$1.04127\pm0.00064$ &$1.04053\pm0.00072$ 
&$1.04097\pm0.00070$ &$1.04039\pm0.00073$ 
& $1.04040\pm0.00074$ &$1.04132\pm0.00063$ &$1.04117\pm0.00063$ &$1.04133\pm0.00064$ \\
$\tau$ 
&$0.089\pm0.013$ 
&$0.097\pm0.015$ 
&$0.089\pm0.013$ 
&$0.096\pm0.015$ 
&$0.096\pm0.014$ 
&$0.089\pm0.013$
&$0.089\pm0.013$&$0.100\pm0.016$\\
$n_s$ 
&$0.9614\pm0.0075$
 &$0.991\pm0.018$ 
&$0.9576\pm0.0088$ 
&$0.985\pm0.019$
&$0.982\pm0.018$ 
&$0.9617\pm0.0073$
&$0.9615\pm0.0074$&$0.9572\pm0.0080$ \\
$log[10^{10} A_s]$ 
&$3.086\pm0.025$ 
&$3.122\pm0.033$ 
&$3.086\pm0.025$ 
&$3.119\pm0.033$ 
&$3.119\pm0.032$
&$3.087\pm0.024$
&$3.149\pm0.026$ &$3.114\pm0.031$ \\
$H_0 [\mathrm{km}/\mathrm{s}/\mathrm{Mpc}]$
& $67.4\pm1.2$ 
&$73.2\pm3.5$ 
&$64.5\pm3.3$ 
&$70.4\pm4.7$ 
&$70.2\pm3.4$
&$84\pm10$
&$67.0\pm1.2$&$67.5\pm1.2$\\
$r$ &$<0.12$ &$<0.19$ &$<0.13$ &$<0.19$ &$<0.18$&$<0.13$&$<0.93$ &$<0.23$\\
$m_{\textrm{a}} (\mu eV)$
 &$81.5\pm1.6$ 
&$76.8\pm2.8$ 
&$81.0\pm1.6$ 
&$77.0\pm2.7$ 
&$77.1\pm2.9$
&$81.6\pm1.5$
&$80.8\pm1.7$ &$81.3\pm1.6$\\
$N_{\textrm{eff}}$ &$(3.046)$ &$3.79\pm0.41$ &$(3.046)$ &$3.71\pm0.41$ &$3.72\pm0.37$&$(3.046)$ &$(3.046)$ &$(3.046)$\\
$\sum m_{\nu} (eV)$ &$(0.06)$ &$(0.06)$ &$<0.97$ &$<0.83$ &$(0.06)$ &$(0.06)$&$(0.06)$&$(0.06)$\\
$w$ &$(-1)$ &$(-1)$ &$(-1)$ &$(-1)$ &$(-1)$&$-1.50\pm0.31$ &$(-1)$
&$(-1)$\\
$m^\textrm{eff}_s (eV)$ &$(0)$ &$(0)$ &$(0)$ &$(0)$ &$<0.87$&$<(0)$ &$(0)$ &$(0)$ \\
$n_t$ &$(0)$ &$(0)$ &$(0)$ &$(0)$ &$(0)$&$(0)$ &$2.19\pm0.87$&$(0)$\\
$d n_s/d\ln k$ &$(0)$ &$(0)$ &$(0)$ &$(0)$ &$(0)$&$(0)$ &$(0)$&$-0.022\pm0.011$\\

\hline\hline
 \end{tabular}}
 \caption{Constraints at $68 \%$ confidence level on cosmological
   parameters from our analysis for Planck+WP, except for the upper
   bounds on the neutrino mass and on the tensor-to-scalar ratio, which refer to $95\%$~CL upper limits.}
 \label{tab0}
 \end{center}
 \end{table*}

\begin{table*}[th]
\begin{center}
\resizebox {19cm}{!}{
\begin{tabular}{|c|c|c|c|c|c|c|c|c|}
\hline\hline
Parameter & ADM+r & ADM+r & ADM+r & ADM+r & ADM+r & ADM+r  & ADM+r  & ADM+r  \\
&&+$N_{\textrm{eff}}$&+$\sum m_{\nu}$&+$\sum
m_{\nu}$+$N_{\textrm{eff}}$&+ $m^\textrm{eff}_s+
N_{\textrm{eff}}$  & +$w$ &+ $n_t$&  + $d n_s/d\ln k$\\
\hline
$\Omega_{\textrm{b}}h^2$ 
&$0.02202\pm0.00028$ 
& $0.02285\pm0.00043$ 
&$0.02193\pm0.00032$ 
&$0.02276\pm0.00046$ 
& $0.02272\pm0.0043$
&$0.02207\pm0.00028$
&$0.02202\pm0.00029$ 
&$0.02234\pm0.00031$\\
$\Omega_{\textrm{a}}h^2$ & 
$0.1186\pm0.0026$ 
& $0.1313\pm0.0057$
 & $0.1191\pm0.0028$ 
& $0.1312\pm0.0059$ 
&$0.1257\pm0.0015$
& $0.1183\pm0.0025$
& $0.1192\pm0.0026$ & $0.1193\pm0.0027$ \\
$\theta$ 
&$1.04138\pm0.00063$ 
& $1.04032\pm0.00071$ 
& $1.04118\pm0.00067$ 
& $1.04023\pm0.00073$ 
&$1.04020\pm0.00075$
& $1.04143\pm0.00062$
&$1.04129\pm0.00065$&$1.04141\pm0.00064$\\
$\tau$ 
&$0.089\pm0.013$ 
& $0.101\pm0.015$ 
& $0.090\pm0.013$ 
&$0.101\pm0.015$ 
& $0.104\pm0.016$
&$0.090\pm0.013$
&$0.089\pm0.013$&$0.104\pm0.016$\\
$n_s$ 
&$0.9649\pm0.0074$ 
& $1.0057\pm0.0173$ 
&$0.9628\pm0.0083$ 
&$1.0032\pm0.0184$
&$1.004\pm0.0175$
& $0.9654\pm0.0073$ 
&$0.9611\pm0.0074$ &$0.1004\pm0.0150$\\
$log[10^{10} A_s]$ 
&$3.084\pm0.025$ 
& $3.136\pm0.034$ 
&$3.085\pm0.025$ 
&$3.135\pm0.034$ 
&$3.134\pm0.033$
&$3.085\pm0.025$
&$3.149\pm0.025$  &$3.123\pm0.031$ \\
$H_0 [\mathrm{km}/\mathrm{s}/\mathrm{Mpc}]$
& $67.7\pm1.2$ 
&$76.0\pm3.6$ 
&$65.9\pm2.8$
 &$74.5\pm4.3$ 
& $73.6\pm3.9$
& $87.1\pm9.1$ 
& $67.5\pm1.2$& $67.7\pm1.2$ \\
$r$ 
&$0.166\pm0.036$ 
& $0.180\pm0.037$ 
&$0.168\pm0.035$
&$0.183\pm0.038$
&$0.183\pm0.038$ 
&$0.168\pm0.035$ 
& $0.172\pm0.047$& 
$0.194\pm0.040$\\
$m_{\textrm{a}} (\mu eV)$ 
&$82.0\pm1.5$ 
& $75.3\pm2.8$ 
&$81.6\pm1.6$ 
&$75.3\pm2.8$
 &$75.3\pm2.9$
&$82.1\pm1.5$
&$81.6\pm1.5$
&$81.5\pm1.6$\\
$N_{\textrm{eff}}$ &$(3.046)$ &$4.13\pm0.43$ &$(3.046)$ &$4.08.\pm0.44$ &$4.08\pm0.42$&$(3.046)$&$(3.046)$&$(3.046)$\\
$\sum m_{\nu} (eV)$ &$(0.06)$ &$(0.06)$ &$<0.78$ &$<0.58$ &$(0.06)$ &$(0.06)$ &$(0.06)$ &$(0.06)$\\
$w$ &$(-1)$ &$(-1)$ &$(-1)$ &$(-1)$ &$(-1)$ &$-1.57\pm0.26$  &$(-1)$ &$(-1)$ \\
$m^\textrm{eff}_s(eV)$&$(0)$ &$(0)$ &$(0)$ &$(0)$ & $<0.63$&$<(0)$ &$(0)$ &$(0)$\\
$n_t$ &$(0)$ &$(0)$ &$(0)$ &$(0)$ &$(0)$ &$(0)$ &$1.66\pm0.51$&$(0)$\\
$d n_s/d\ln k$ &$(0)$ &$(0)$ &$(0)$ &$(0)$ &$(0)$ &$(0)$ & $(0)$& $-0.0278\pm0.0099$\\

\hline\hline
 \end{tabular}}

 \caption{Constraints at $68 \%$ confidence level on cosmological
   parameters from our analysis for Planck+WP+BICEP2, except for the
   bounds on the neutrino mass, which refer to $95\%$~CL upper limits.}
 \label{tab1}
 \end{center}
 \end{table*}

Here we present the results for the allowed axion mass ranges in the
scenario which would survive once the BICEP2 findings concerning the
tensor to scalar ratio and, consequently, the energy scale associated
to inflation, are confirmed by ongoing and near future searches of
primordial B modes. In this case, the PQ symmetry should be broken
after inflation. We shall restrict ourselves to such scenario  in the following. 

Tables~\ref{tab0} and \ref{tab1} depict the results for the different cosmologies explored in
this study, for two possible data combinations: \textit{(a)} Planck
temperature data + WMAP polarization (WP)
 and \textit{(b)} Planck temperature data, WP and
BICEP2 measurements. The constraints on the tensor to scalar ratio are quoted for a reference scale
 of $k_0=0.05$~Mpc$^{-1}$. For the sake of simplicity, we do not show
 here all the results with the BAO measurements included
 in the numerical analyses. We shall
 quote the values of the cosmological parameters resulting from the
 analyses with BAO data included only in the cases in which these values notably differ from the
 results obtained without considering BAO measurements.

In the standard ADM model we find  $m_{\textrm{a}}=81.5\pm1.6\,\mu$eV ($m_{\textrm{a}}=82.0\pm1.5\,\mu$eV)
from Planck, WP (+BICEP2) data, corresponding to a cold dark matter
energy density of $\Omega_{\textrm{a}} h^2=0.1194\pm 0.0027$
($\Omega_{\textrm{a}} h^2= 0.1186\pm 0.026$). When adding BAO measurements, the
former value translates into
$m_{\textrm{a}}=82.2\pm1.0\,\mu$eV. Therefore, the inclusion of BAO data
 reduces mildly the error on $m_{\textrm{a}}$.

\begin{figure*}
\begin{tabular}{c c}
\includegraphics[width=8cm]{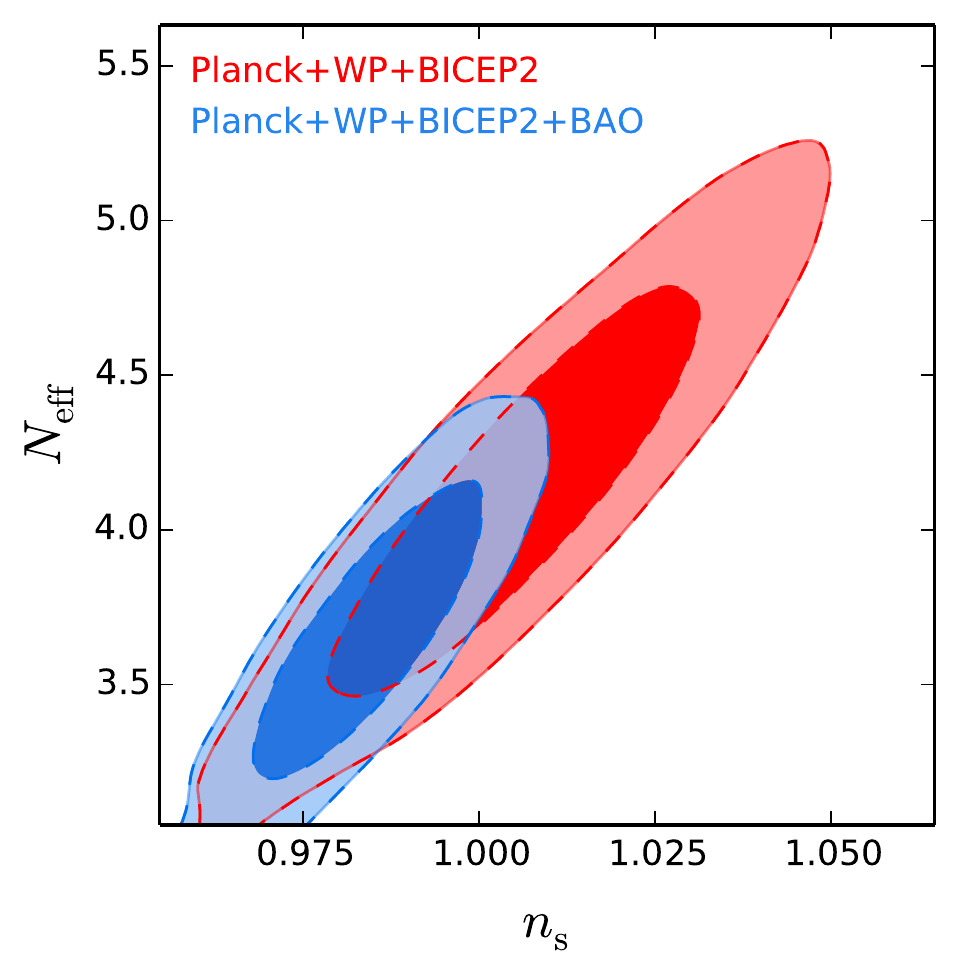}&\includegraphics[width=8.2cm]{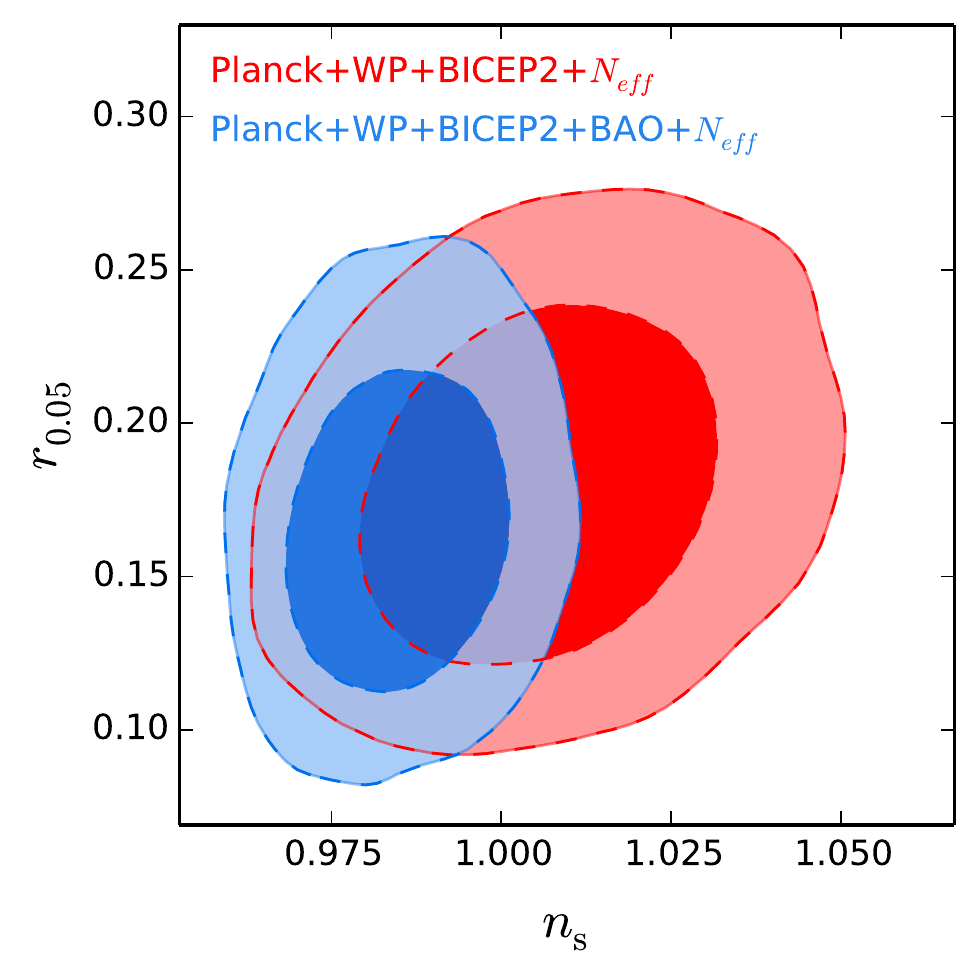}\\
\end{tabular}
 \caption{Left panel: the red contours show the $68\%$ and $95\%$~CL allowed
  regions from the combination of Planck data, WP and BICEP2
  measurements in the ($n_s$, $\neff$) plane. The blue contours depict the
  constraints after the BAO data sets are added in the analysis. Right
  panel: as in the left panel but in the ($n_s$, $r$) plane.}
\label{fig:fignnu}
\end{figure*}

When allowing $\neff$ to be a free parameter to extend the minimal ADM scenario
to scenarios in which additional relativistic species are present, we
find that $m_{\textrm{a}}=76.8\pm2.8\mu$eV and $\neff=3.79\pm 0.41$ for Planck+WP,
and $m_{\textrm{a}}=75.3\pm2.8\mu$eV and $\neff=4.13\pm 0.43$
after combining Planck data with WP and BICEP2 measurements. 
When BAO data sets are included in the analysis, the former values are
translated into $m_{\textrm{a}}=76.6\pm 2.6\mu$eV and $\neff=3.69\pm
0.30$. Therefore, there exists from CMB data a $2- 3 \sigma$evidence for extra
radiation species. The higher value of $\neff$ found when
considering tensor modes and BICEP2 simultaneously was first found in
Ref.~\cite{Giusarma:2014zza}, where it was also pointed out that the tension between
the tensor-to scalar ratio $r$ extracted by Planck and WP data and
the value of $r$ found by BICEP2 data is less evident when 
$\neff>3$. The reason for this is because, if the value of $\neff>3$,
the power in the CMB damping tail is suppressed. This can be
compensated by a higher scalar spectral index $n_s$ which in turn,
will reduce the power at large scales. This power reduction at small
multipoles can be compensated by increasing the tensor to scalar ratio $r$, and the 
overall result is a positive correlation between $\neff$ and $r$. 
This effect is illustrated in Fig.~\ref{fig:fignnu}, where the left
panel depicts the strong positive correlation between $\neff$ and
$n_s$ and the left panel shows the relation between $n_s$ and
$r$. Concerning exclusively  CMB
data,  a larger value of $\neff$ can be compensated with a larger
value of $r$ (and viceversa), being the degeneracy among these two parameters
 mildly broken when considering as well BAO data in the MCMC analysis. Thus the preference for $\neff > 3$, already present in the Planck data,
is further increased by the inclusion of the BICEP2 likelihood that assigns a large probability
to the $r\simeq 0.2 $ region.

The larger value of $\neff$ results in a smaller axion mass (and in a larger associated error)
due to the well-known existing correlation between $\neff$ and
$\Omega_{\textrm{a}} h^2$ (that is, the cold dark matter energy
density) when considering
only CMB data, since it is possible to increase both to leave the
redshift of matter-radiation equality unchanged.
This effect can be clearly noticed from the results depicted in
Tabs.~\ref{tab0} and \ref{tab1}, where the value of $\Omega_{\textrm{a}} h^2$ is about
$\sim 2\sigma$ larger than the value found in
the minimal scenario with no extra dark radiation species.
The error on the $\Omega_{\textrm{a}} h^2$ cosmological parameter is also larger.  Given that $\Omega_{\textrm{a}} h^2$ is inversely proportional to $m_{\textrm{a}}$, this results in an anticorrelation
between $\neff$ and $m_{\textrm{a}}$. The large degeneracy between $\neff$ and $\Omega_{\textrm{c}} h^2\equiv\Omega_{\textrm{a}}h^2$
also drives the large value of $H_0$ found in this case. The degeneracy is partly broken by the
inclusion of BAO information:  when the BAO data sets are considered,
both $H_0$ and $\neff$ are closer to their ADM+$r$ values, being
$H_0=71.7\pm 1.9$ and $\neff=3.69\pm 0.30$ respectively.

We also consider a model in which the active neutrino mass is a free
parameter. In this case, $m_{\textrm{a}}=81.6\pm1.6\mu$eV 
after combining Planck data with WP and BICEP2 measurements, while
$m_{\textrm{a}}=82.4\pm1.1\mu$eV when BAO
data sets are also considered. However, in this 
$\Lambda$CDM plus massive neutrino scenario, the neutrino
mass bounds are unaffected when considering tensors and BICEP2
data. Indeed, the $95\%$~CL bound on the total neutrino mass we get after considering 
all the data explored in this paper, $\sum m_\nu<0.25$~eV, agrees
perfectly with the one found when neither tensors nor BICEP2
data are included in the analyses~\cite{Giusarma:2014zza}. We have also explored here the case 
in which the three massive active neutrinos
coexist with $\Delta \neff$ massless species. The numerical results
without BAO data are presented in the fifth column of Tabs.~\ref{tab0} and
\ref{tab1}. The values
 obtained for the axion mass and for the number of relativistic degrees of
freedom in this scenario are very close to the ones reported above for
the $\neff$ cosmology, finding, from CMB data, evidence for extra
dark radiation species at more than $2\sigma$. When considering the
full data set exploited here, including BAO measurements, the bound on the neutrino
mass becomes less stringent than in the three massive
neutrino scenario due to the strong $\sum m_\nu$-$\neff$
degeneracy: we find a $95\%$~CL bound of $\sum m_\nu < 0.47$~eV from
the combination of Planck data with WP, BICEP2 and BAO measurements.
Notice that, as in the case of the ADM plus $\neff$ relativistic degrees of
freedom model, the mass of the axion is smaller and the value of the
Hubble constant is larger than in the standard ADM scenario. The
reason for that is due to the large existing degeneracy between
$\Omega_{\textrm{a}} h^2$ and $\neff$ when considering CMB data
only: notice the higher value of $\Omega_{\textrm{a}} h^2$ in
Tabs.~\ref{tab0} and \ref{tab1}, when compared to its value in the
standard ADM+$r$ scenario.

The last neutrino scenario we analyse here is the case in which there
are $\Delta \neff$ sterile massive neutrino species, characterised by a mass
$m^\textrm{eff}_s$, which, for instance, in the case of a thermally-distributed
sterile neutrino state, reads
\begin{equation}
\label{parameter}
m^\textrm{eff}_s= (T_s/T_\nu)^3m_s=(\Delta \neff)^{3/4} m_s~,
\end{equation}
where $T_s$, $T_\nu$ are the current temperature of the sterile and active
neutrino species, respectively, and $m_s$ is the true sterile neutrino
mass. We recall however that the parameterization in terms of $\Delta \neff$ 
and $m^\textrm{eff}_s$ is more general, and also includes, among others,
the case of a Dodelson-Widrow sterile neutrino (in which case $m^\textrm{eff}_s= \Delta\neff\, m_s$).
 For this particular case we have fixed
the mass of the three light neutrino species $\sum m_\nu=0.06$~eV,
i.e. the minimum value indicated by neutrino oscillation data. In this
case, we find an axion mass, a number of neutrino species and a
effective sterile neutrino mass  of $m_{\textrm{a}}=75.3\pm2.9\mu$eV,
$\neff=4.08\pm 0.42$ and $m^\textrm{eff}_s<0.63$~eV at $95\%$~CL
($m_{\textrm{a}}=76.5\pm2.6\mu$eV, $\neff=3.82\pm 0.32$ and
$m^\textrm{eff}_s<0.51$~eV at $95\%$~CL) before (after) the combination of Planck, WP and BICEP2
measurements with BAO results. As
previously explained and as expected, the mean values for $\neff$ are
considerably larger than those found in the absence of BICEP2
data. Concerning the bounds on the effective sterile neutrino mass,
the values are mildly shifted when the BICEP2
measurements are addressed due to the anticorrelation between $\neff$
and $m^\textrm{eff}_s$, being the $95\%$~CL constraints on the
neutrino mass constraints tighter when considering BICEP2 data. Our
findings agree with the recent results presented in Refs.~\cite{Zhang:2014dxk,Dvorkin:2014lea,Archidiacono:2014apa}, which also include BICEP2 data.
Note that the mean value of the cold dark matter density, made by
axions, is, again, larger than what is found in the standard ADM+$r$ scenario.

The next scenario explored here is a $w$CDM model with a free, constant, dark energy
equation-of-state parameter $w$. Both the values of the axion masses and the
value of the tensor to scalar ratio $r$ are very close to their values
in the ADM model. However, when the 
BAO data are not considered, the equation-of-state parameter is
different from $-1$ at $\sim 95\%$ CL ($w=-1.57\pm 0.26$),
and we also find a very large value for $H_0=87.1 \pm 9.1$~km/s/Mpc.
 The addition of BAO constraints make both the value
of the Hubble constant $H_0$ and of the
dark energy equation of state $w$ much closer to their expected values
within a minimal $\Lambda$CDM scenario, being the values of these two parameters 
 $w=-1.12\pm 0.12$ and $H_0=70.5 \pm 2.8$~km/s/Mpc, respectively. This illustrates the
highly successful constraining power of BAO data concerning dark energy measurements.

\begin{figure*}
\begin{tabular}{c c}
\includegraphics[width=8cm]{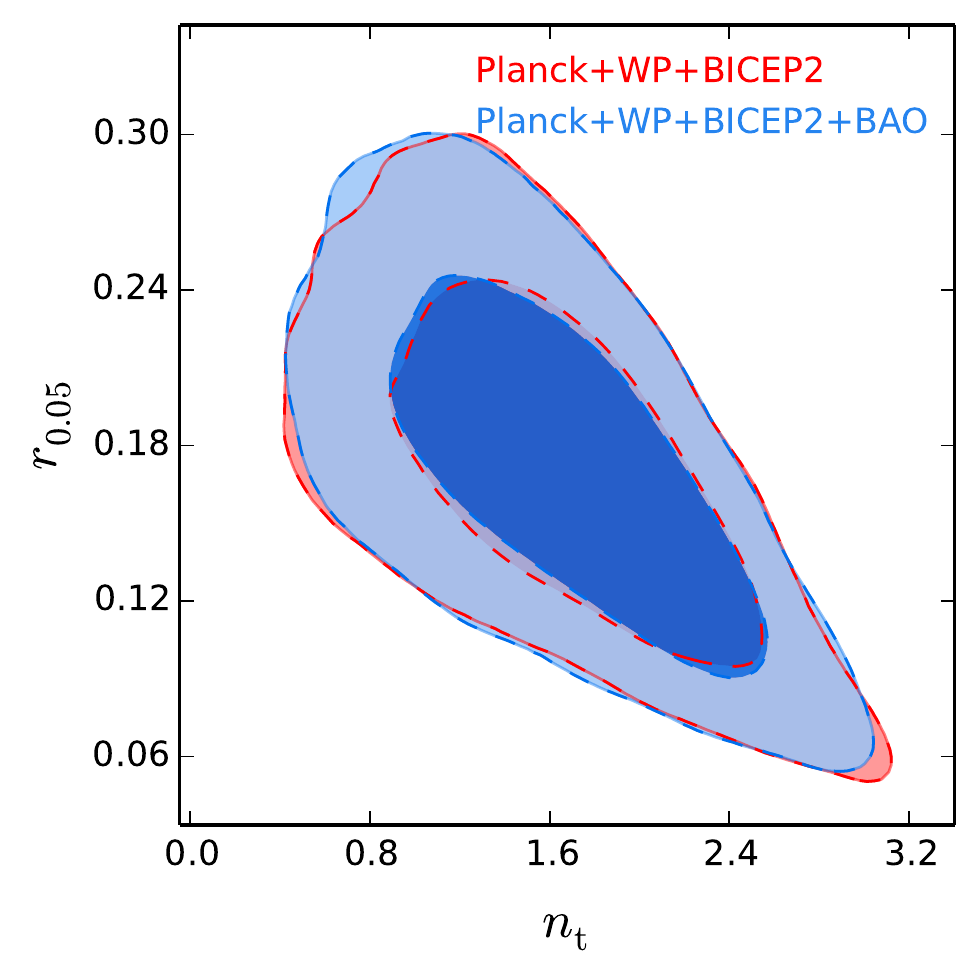}&\includegraphics[width=8.cm]{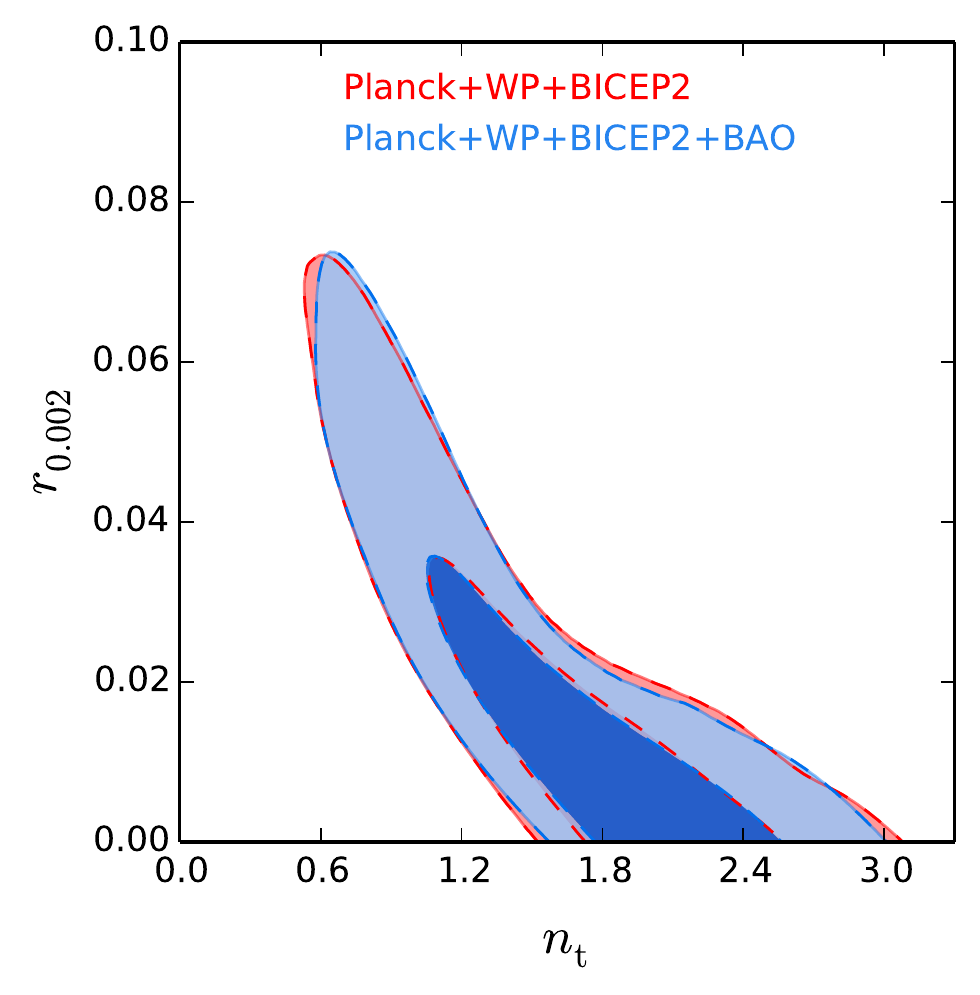}\\
\end{tabular}
 \caption{Left panel: the red contours show the $68\%$ and $95\%$~CL allowed
  regions from the combination of Planck data, WP and BICEP2
  measurements in the ($n_t$, $r_{0.05}$) plane, referring these
  limits to a scale of $k_0=0.05$~Mpc$^{-1}$. The blue contours depict the
  constraints after the BAO data sets are added in the analysis. Right
  panel: as in the left panel but for a scale $k_0=0.002$~Mpc$^{-1}$.}
\label{fig:nt}
\end{figure*}

Very recently, the authors of Ref.~\cite{Gerbino:2014eqa}  have
extracted the tensor spectral index from the BICEP2
measurements. The standard inflationary paradigm predicts a small,
negative, tensor spectral index. More concretely, the inflation
consistency relation implies that $n_T \simeq -r/8$. We shall relax this constrain here,
leaving $n_T$ as a free parameter.  We rule out a scale invariant
tensor spectrum with $3\sigma$ significance when considering CMB data
only. The addition of BAO measurements does not change significantly
these results,  see Fig. ~\ref{fig:nt}. As expected, the axion mass constraints
are unaffected by the presence of a free $n_T$. The
value of the tensor-to-scalar ratio we find is $r = 0.172 \pm 0.047$
using the Planck+WP+BICEP2 dataset.  The fact that the data support a non-zero spectral index for tensors
also implies that $r$ strongly depends on the scale $k_0$. The
corresponding $95\%$~CL limit on
$r_{0.002} \equiv r(k = 0.002\,\Mpc^{-1})$ is $r_{0.002} <0.055$ for
the Planck+WP+BICEP2 datasets, see Fig. ~\ref{fig:nt}, right panel.

\begin{figure*}
\begin{tabular}{c c}
\includegraphics[width=8.6cm]{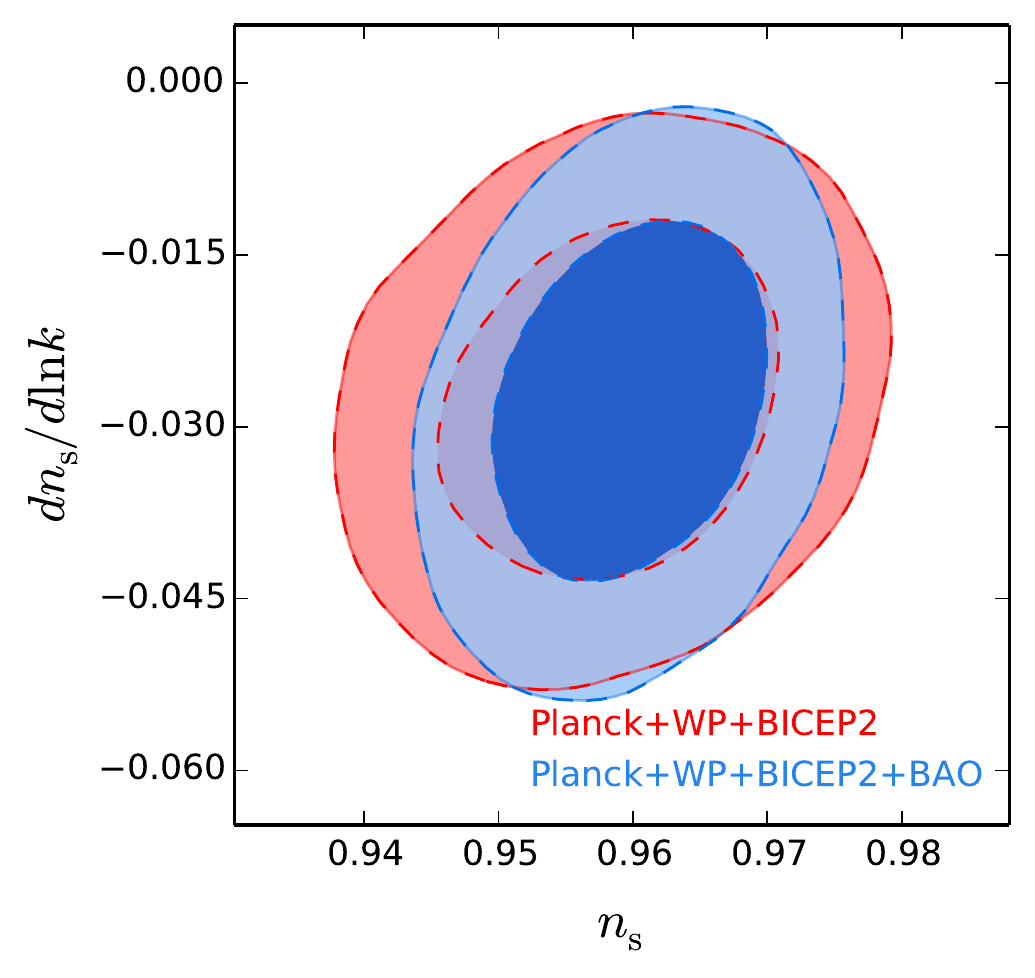}&\includegraphics[width=8.2cm]{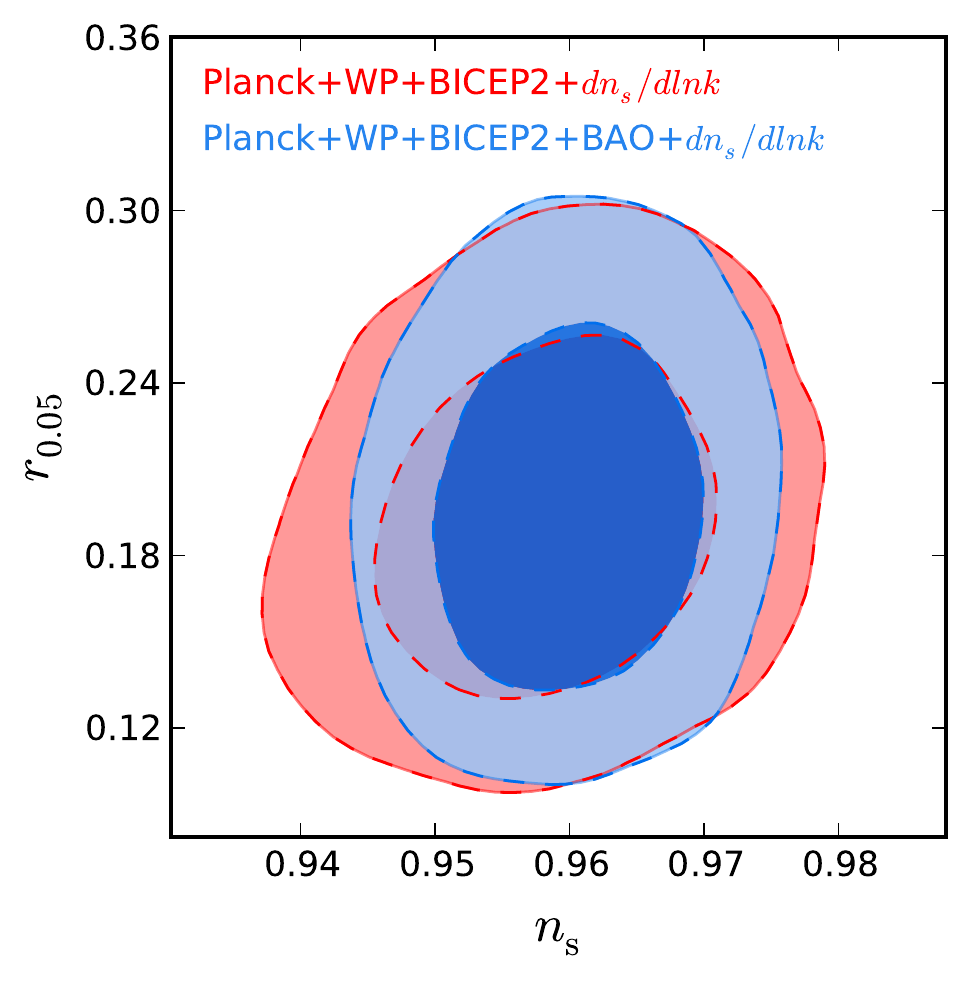}\\
\end{tabular}
 \caption{Left panel: the red contours show the $68\%$ and $95\%$~CL allowed
  regions from the combination of Planck data, WP and BICEP2
  measurements in the ($n_s$, $d n_s/d\ln k$) plane. The blue contours depict the
  constraints after the BAO data sets are added in the analysis. Right
  panel: as in the left panel but in the ($n_s$, $r$) plane.}
\label{fig:fignrun}
\end{figure*}

The latest extended scenario considered is the one with a running of
the scalar spectral index $n_{\textrm{run}}=d n_s/d\ln k$. This
minimal extension was firstly addressed in the context of a
$\Lambda$CDM scenario by the BICEP2 collaboration, in order to relax
the discrepancy between their measurements of the tensor to scalar
ratio $r$ and the limits on the same quantity arising from Planck
data~\cite{Ade:2014xna,Ade:2014gua}.  The reason for that is due to
the degeneracy between the running and the scalar spectral index: a
negative running of the spectral index can be compensated with a larger scalar
spectral index, which will decrease the CMB temperature power spectra
at large scales. This lowering effect at low multipoles can be compensated  
 with a  higher tensor contribution to the temperature fluctuations
 (by increasing $r$). The former degeneracies are depicted in Fig.~\ref{fig:fignrun}.
The BICEP2 collaboration reports $d n_s/d\ln k=-0.022\pm0.010$ at
$68\%$~CL, whose absolute value is smaller than what we find in the context of the ADM
scenario, $d n_s/d\ln k=-0.028\pm 0.010$ at
$68\%$~CL.

\section{Conclusions}
\label{sec:sec4}

The exact nature of dark matter is still an open issue, involving both particle physics and cosmology.
A well-motivated candidate for the role of DM is the axion, the pseudo Nambu-Goldstone boson
associated to the breaking of the PQ symmetry, proposed to explain the absence 
of CP violation in strong interactions. The axion can be created non-thermally in the early Universe
through the misalignment mechanism and the decay of axionic strings,
and its mass is inversely proportional to the scale $f_{\textrm{a}}$ at which the PQ symmetry is broken.
This can happen, in principle, either before or after the end of inflation; however, the
large value of the tensor-to-scalar ratio implied by the recent BICEP2 observations
seem to exclude the first possibility, as it would imply the presence of a large isocurvature component
in the primordial perturbations, far above the current observational limits.

We have presented here the constraints on the ''axion dark matter''
scenario in which the PQ symmetry is broken after inflation,  
using the most precise CMB data available to date (including the recent BICEP2 on the spectrum
of B-modes), as well as the recent and most precise distance BAO constraints to date from the BOSS Data Release 11 
(DR11). We find that, in the minimal ADM scenario, the largest dataset implies $m_{\textrm{a}} = 82.2 \pm 1.1\,\mu\eV$,
corresponding to $f_{\textrm{a}} = (7.54 \pm 0.10) \times 10^{10} \,\GeV$. These values
change to $m_{\textrm{a}} = 76.6 \pm 2.6\,\mu\eV$ and $f_{\textrm{a}} = (8.08 \pm 0.27) \times 10^{10} \,\GeV$
when we consider a model with an additional number of relativistic degrees of freedom $\neff$.
In that case, we also find that a non-standard value for $\neff$ is preferred at more than 95\% CL.
We find similar results for $m_{\textrm{a}}$ and $\neff$ if we also allow the neutrino mass to vary. 
For what concerns the latter parameter, we obtain $\sum m_\nu < 0.25\,\eV$ at 95\% CL for the 
Planck+WP+BICEP2+BAO dataset if we fix $\neff=3.046$; the constraint is degraded
to $\sum m_\nu<0.47\,\eV$ when $\neff$ is allowed to vary, due to the
strong degeneracy between these two parameters. The case of
  $\Delta \neff$ sterile massive neutrino species,  
characterised by a mass $m^\textrm{eff}_s$, has also been analysed and
the constraints on the axion mass are very similar to those previously
quoted. We find  $\neff=4.08\pm 0.42$ and $m^\textrm{eff}_s<0.63$~eV
at $95\%$~CL from the combination of Planck, WP and BICEP2
measurements, finding evidence for extra dark radiation species at more than $2\sigma$.

We have also addressed other extensions to the baseline ADM model,
by considering, one at a time, the dark energy equation-of-state parameter $w$,
the tensor spectral index $n_T$ and the running $n_{\mathrm{run}}$ of the  scalar spectral index.
In none of these extended models the results for $m_{\textrm{a}}$ change significantly. The BICEP2 data, however, drive a preference for non-zero tensor 
spectral index or non-zero scalar running in the corresponding models.

The search for axion dark matter is also the target of laboratory experiments like
the Axion Dark Matter eXperiment (ADMX) \cite{Asztalos:2011bm}, that uses a tunable microwave
cavity positioned in a high magnetic field to detect the conversion of axions into photons.
This is enhanced at a resonant frequency $\nu =m_a/2\pi$; for the typical masses found in our study,
this corresponds to a frequency $\nu\simeq 20$ GHz. ADMX has been operating in the range
$0.3 - 1\,\mathrm{GHz}$, thus being able to exclude DM axions in the mass range between
1.9 and 3.53 $\mu$eV \cite{Asztalos:2003px,Asztalos:2009yp}. ADMX is currently undergoing an upgrade that will extend its frequency
range up to a few GHzs (i.e., masses in the $~ 10\,\ueV$ range) \cite{vanBibber:2013ssa}, which is unfortunately 
still not enough to detect DM axions in the mass range implied by cosmological observations\footnote{This is still true even if one assumes $\alpha_\mathrm{dec}\simeq0$, that corresponds, for a given of $\Omega_\mathrm{a} h^2$,
to the smallest value of the axion mass. We also note that it does not help
either if axions only make up for part of the 
total DM content of the Universe. In fact, since $\Omega_a h^2 \propto m_a^{-6/7}$, having $\Omega_a h^2 < \Omega_c h^2$ would
just shift the resonant frequency to even higher values.},
if the PQ symmetry is broken after inflation (as implied by the recent BICEP2 data). However, a second, smaller experiment called
ADMX-HF is currently being built, that will allow to probe the $4 - 40$ Ghz range \cite{vanBibber:2013ssa}, thus 
being in principle sensitive to axion masses in the $\sim 100\,\ueV$ range, allowing to directly test the ADM scenario,
at least in its simplest implementation.

\section{Acknowledgments}
M.L. is supported by Ministero dell'Istruzione, dell'Universit\`a e
della Ricerca (MIUR) through the PRIN grant \emph{Galactic and extragalactic polarized microwave
emission'} (contract number PRIN 2009XZ54H2-002). Part of this work was carried out
while M.L. was visiting the Instituto de F\'isica Corpuscular in Valencia, whose hospitality is
kindly acknowledged, supported by the grant \emph{Giovani ricercatori} of the University of Ferrara,
financed through the funds \emph{Fondi 5x1000 Anno 2010} and \emph{Fondi Unicredit
2013}. O.M. is supported by the Consolider Ingenio project CSD2007-00060, by
PROMETEO/2009/116, by the Spanish Ministry Science project FPA2011-29678 and by the ITN Invisibles PITN-GA-2011-289442.




\end{document}